\documentclass[twocolumn,prl,superscriptaddress,amsmath,amssymb,longbibliography]{revtex4-1}

\usepackage{graphicx}

\usepackage{hyperref}

\usepackage{color}

\usepackage[dvipsnames]{xcolor}

\usepackage{upgreek}

\usepackage{float}

\usepackage{upgreek}

\usepackage{amsmath,amssymb}

\newcommand{\degree}{\ensuremath{^{\circ}}}

\newcommand{\micro}{\ensuremath{\upmu}}

\newcommand{\ket}[1]{\ensuremath{\left| #1 \right\rangle}}

\begin{document}

\title{Enhanced widefield quantum sensing with nitrogen-vacancy ensembles using diamond nanopillar arrays}

\author{D. J. McCloskey}

\affiliation{School of Physics, University of Melbourne, Parkville, VIC 3010, Australia}

\author{N. Dontschuk}

\affiliation{School of Physics, University of Melbourne, Parkville, VIC 3010, Australia}

\affiliation{Centre for Quantum Computation and Communication Technology, School of Physics, University of Melbourne, Parkville, VIC 3010, Australia}

\author{D. A. Broadway}

\affiliation{School of Physics, University of Melbourne, Parkville, VIC 3010, Australia}

\affiliation{Centre for Quantum Computation and Communication Technology, School of Physics, University of Melbourne, Parkville, VIC 3010, Australia}
 
\author{A. Nadarajah}

\affiliation{School of Physics, University of Melbourne, Parkville, VIC 3010, Australia}

\author{A. Stacey}

\affiliation{Centre for Quantum Computation and Communication Technology, School of Physics, University of Melbourne, Parkville, VIC 3010, Australia}

\affiliation{Melbourne Centre for Nanofabrication, Clayton, VIC 3168, Australia}

\author{J.-P. Tetienne}

\affiliation{School of Physics, University of Melbourne, Parkville, VIC 3010, Australia}

\author{L. C. L. Hollenberg}

\affiliation{School of Physics, University of Melbourne, Parkville, VIC 3010, Australia}

\affiliation{Centre for Quantum Computation and Communication Technology, School of Physics, University of Melbourne, Parkville, VIC 3010, Australia}

\author{S. Prawer}

\affiliation{School of Physics, University of Melbourne, Parkville, VIC 3010, Australia}

\author{D. A. Simpson}

\affiliation{School of Physics, University of Melbourne, Parkville, VIC 3010, Australia}
 
\begin{abstract}

\noindent Quantum sensors based on optically active defects in diamond such as the nitrogen vacancy (NV) centre represent a promising platform for nanoscale sensing and imaging of magnetic, electric, temperature and strain fields. Enhancing the optical interface to such defects is key to improving the measurement sensitivity of these systems. Photonic nanostructures are often employed in the single emitter regime for this purpose, but their applicability to widefield sensing with NV ensembles remains largely unexplored. Here we fabricate and characterize closely-packed arrays of diamond nanopillars, each hosting its own dense, near-surface ensemble of NV centres. We explore the optimal geometry for diamond nanopillars hosting NV ensembles and realise enhanced spin and photoluminescence properties which lead to increased measurement sensitivities (greater than a factor of 3) when compared to unpatterned surfaces. Utilising the increased measurement sensitivity, we image the mechanical stress tensor in each nanopillar across the arrays and show the fabrication process has negligible impact on in-built stress compared to the unpatterned surface. Our results demonstrate that photonic nanostructuring of the diamond surface is a viable strategy for increasing the sensitivity of ensemble-based widefield sensing and imaging.

\end{abstract}
\maketitle
\section*{Introduction}
\begin{figure*}
\includegraphics[width = \textwidth]{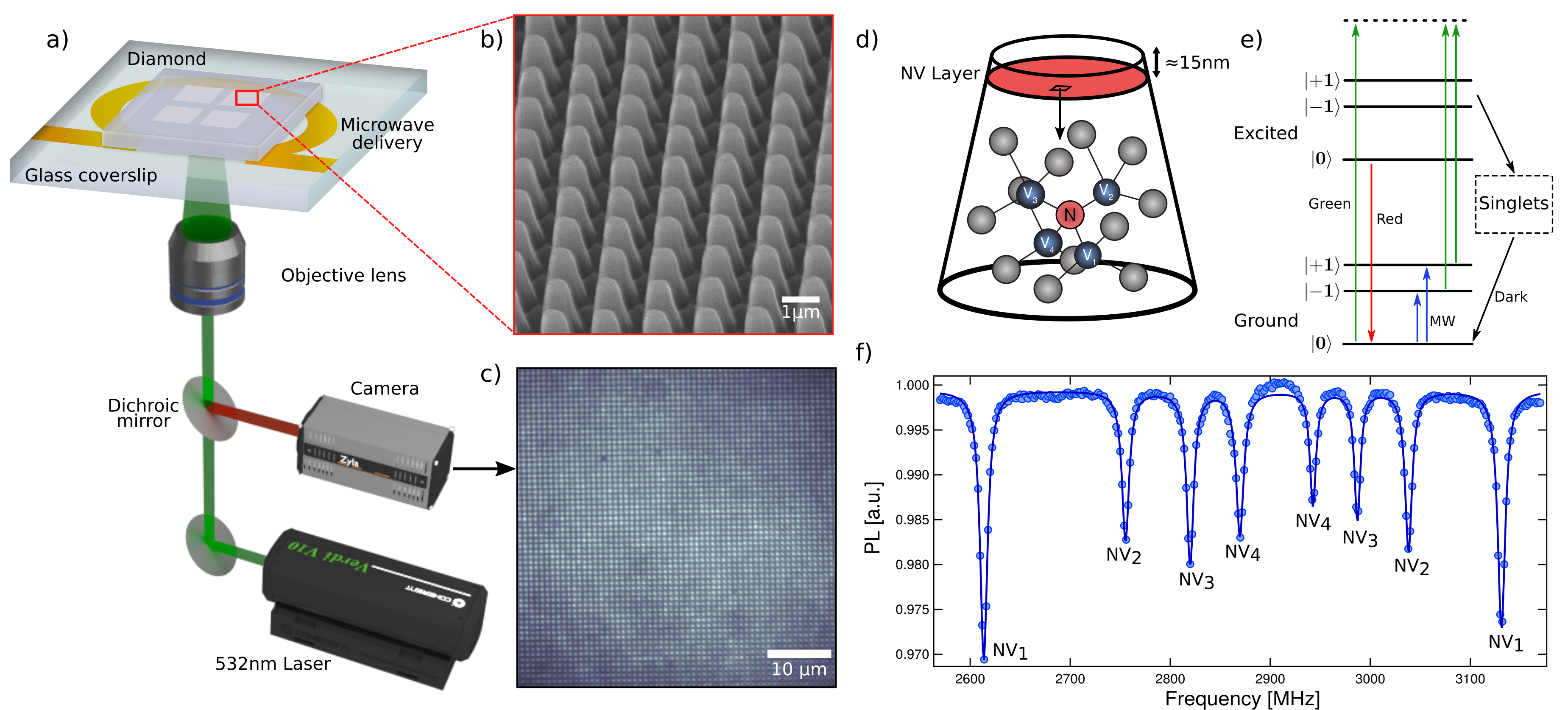}
\caption{\textbf{Widefield sensing with arrays of nanopillar hosted NV ensembles.} a) Widefield measurement setup, consisting of 532nm laser excitation onto a diamond sample, and collection of NV emissions into an sCMOS camera. Microwave (MW) excitation is provided to the NV centres for magnetic resonance measurements through an omega-shaped gold loop patterned onto a glass coverslip. b) Scanning-electron micrograph of a section of a 500nm tip-diameter pillar array. c) Widefield PL image of a section of a 600\,nm tip-diameter pillar array. d) Schematic image of a single diamond nanopillar, which contains all four possible crystallographic orientations of NV centres and funnels their emissions through the diamond substrate and into collection optics. e) Simplified energy-level diagram of an NV centre. A 532\,nm laser non-resonantly excites the NV from the ground to the excited state via a spin-conserving transition, while microwave-frequency excitation induces transitions into the magnetic sublevels of the ground state. Photoluminescence emission of the NV centre is reduced in these sublevels due to an increased probability of relaxation through a non-fluorescent (spin-singlet) decay pathway. f) Optically-detected magnetic-resonance signal obtained from a single 600\,nm pillar with a magnetic field applied out of alignment with the <111> axes of the sample. The four pairs of peaks correspond to the four orientations the NV centre can assume in the diamond lattice, with the lower and higher frequency peaks being transitions into the $\ket{-1}$ and $\ket{+1}$ sublevels respectively. Circles are measured data points, while the solid line is a fit comprised of eight Lorentzian peaks.}
\label{overall_diagram}
\end{figure*}

\noindent Nitrogen-vacancy (NV) defect centres in diamond are a promising system for engineering a variety of sensors owing to their spin-dependent fluorescence and room-temperature quantum coherence properties \cite{8911201520130701, doi:10.1146/annurev-physchem-040513-103659}. Single NV centres have been used to demonstrate high-resolution sensing of a variety of quantities including magnetic and electric fields, mechanical strain, and temperature \cite{PhysRevLett.112.047601, 0034-4885-77-5-056503, Casola2018, Dolde2011, Kucsko2013}. Meanwhile, ensembles of NV centres have been employed to extend this nanoscale sensing functionality to the regime of widefield microscopy (fig. \ref{overall_diagram}\,a), enabling real-space studies of electronic devices and biological systems \cite{1367-2630-13-4-045021, LeSage2013, doi:10.1063/1.3385689, Simpson2016, Tetiennee1602429, Broadway2018}.\newline

\noindent A common method applied to increase the measurement sensitivity of NV centres is to increase the photon collection efficiency, which is typically less than 5\% for NVs contained in bulk diamond \cite{4837081120100301}. This approach has been extensively utilised for applications with single NV centres via photonic structuring of the diamond surface \cite{2543809120150114,doi:10.1063/1.4902818,4837081120100301,doi:10.1063/1.4952953, doi:10.1063/1.3519847, doi:10.1063/1.3573870, PhysRevLett.114.136402, PhysRevX.5.031009, doi:10.1063/1.4871580}. In addition, patterning of diamond surfaces to increase the surface-to-volume ratio has recently been shown to improve the sensitivity of area-averaged NMR measurements \cite{2877528020170804}, however the prospects for enhancing the performance of widefield imaging systems remain largely unexplored. In particular, optimal surface geometries for imaging with near-surface ensembles of NV centres have yet to be identified.\newline

\noindent Here we consider a new approach for widefield imaging with NV centres, coupling near-surface ensembles to closely-packed arrays of diamond nanopillars for enhanced sensitivity. We report the detailed characterisation of discrete ensembles of near-surface NV centres contained within the tips of diamond nanopillars (figure \ref{overall_diagram}\,b-d) with diameters varying from 400\,nm to 1\,\micro m. Our results demonstrate that such nanostructuring enhances both the spin and fluorescence characteristics of the ensembles by comparing to unstructured regions on the same sample. We find that pillars increase the rate of photon collection by more than an order of magnitude without sacrificing any spin coherence properties, yielding a sensitivity increase of more than 3 times. To illustrate the utility of this system for widefield microscopy, we perform optically-detected magnetic resonance (ODMR) spectral imaging (figure \ref{overall_diagram}\,e, f) and reconstruct the mechanical stress tensor in the pillars with single-pillar (sub-micron) resolution. \newline

\begin{figure}
\includegraphics[width=\columnwidth]{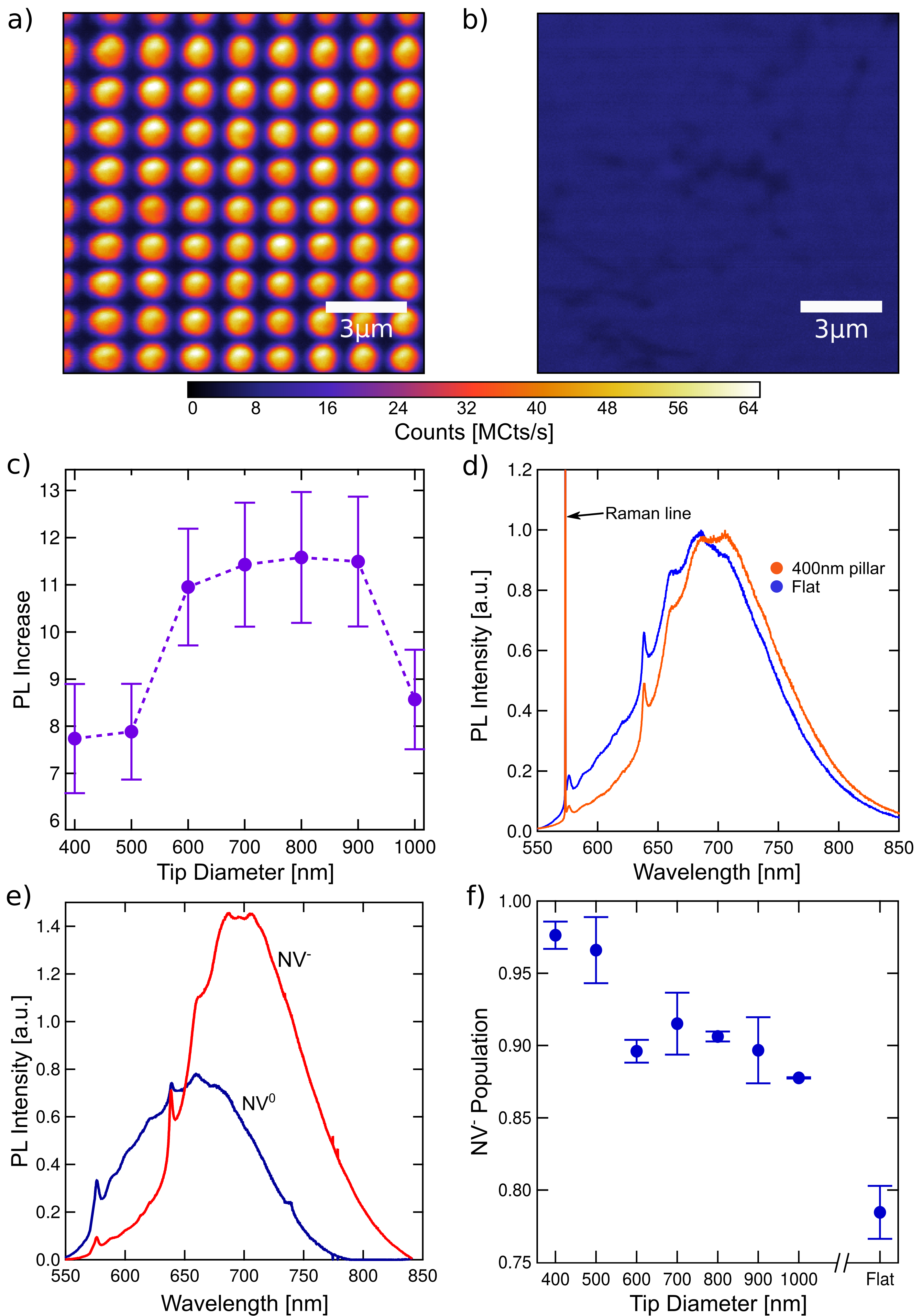}
\caption{\textbf{Photoluminescence characterisation of diamond nanopillars.} a) Scanning confocal photoluminescence map of an array of 800\,nm pillars and b) a flat surface on the same sample. Count rates have been corrected for the 10$\times$ neutral density filter placed in front of the APD. c) PL count rate increase over the flat surface for individual pillars. Error bars are derived from standard deviations of measured count rates. d) PL spectra obtained for the flat surface and from a section of 400\,nm tip diameter pillars. e) Estimated NV\textsuperscript{-} and NV\textsuperscript{0} component spectra extracted from measured spectra using a non-negative matrix factorization method. f) Estimated fraction of negatively-charged NV centres as a function of pillar tip diameter as computed using the spectra in panel (e). Error bars indicate the maximum and minimum population ratios obtained for each diameter.}
\label{PL_enhancement}
\end{figure}

\section*{Fabrication and Photoluminescence Characterisation}

\noindent To quantify improvements in sensitivity, we utilize the general scaling factor for the sensitivity of an NV spin-readout measurement in the photon shot-noise limited regime \cite{Taylor2008}

\begin{equation}
\eta \propto \frac{1}{\gamma C\sqrt{\epsilon N_{NV}  T}},
\end{equation}


\noindent where $\gamma$ is the gyromagnetic ratio of the NV centre, C is the spin contrast,  $\epsilon$ is the rate of photon collection per NV centre, $N_{NV}$ is the number of NV centres in the collection volume, and $T$ is the relevant decoherence time of the NV centres under a given sensing protocol: $T\textsubscript{2}^*$ (intrinsic decoherence) for static field measurements, $T_{2}$ (decoherence under a dynamical decoupling sequence) for measurements of oscillatory signals \cite{Taylor2008,Maze2008}.\newline

\noindent We use a 2\,mm$\times$2\,mm$\times$150\,\micro m $<$100$>$ electronic-grade single-crystal diamond wafer synthesized by chemical vapor deposition (Element 6) and implanted with 10\textsuperscript{13}cm\textsuperscript{-2} \textsuperscript{15}N ions at 6\,keV (InnovIon) and annealed in vacuum at 950\degree C for 3 hours. Arrays of pillars 1.7\,\micro m high were fabricated across areas 100\,\micro m $\times$ 100\,\micro m in size using electron-beam lithography and an oxygen reactive-ion etching process, with pillar tip diameters ranging from 400\,nm to 1000\,nm (steps of 100\,nm) and a pillar-to-pillar pitch equal to twice the tip diameter in all cases. Additional 300\,\micro m\textsuperscript{2} areas were masked from etching to preserve unpatterned regions on the sample to act as experimental controls.\newline

\noindent Quantification of PL collection enhancement was carried out using a custom-built scanning confocal microscope utilising a 60$\times$ 1.40\,N.A. oil-immersion lens, resulting in a diffraction-limited spot size of 400\,nm for the 532\,nm CW laser used to excite the NVs. For confocal microscopy measurements, 55\micro W of excitation power was used to pump the NV centres and PL was measured by passing the collected emission through a 564nm long-pass dichroic mirror followed by a 690nm band-pass filter (width of 120nm) and a 10x neutral density (ND) filter into an avalanche photodiode (APD) \cite{PhysRevB.94.155402}.\newline

\noindent To determine the PL enhancement offered by our arrays over the unpatterned surface (figure \ref{PL_enhancement} a, b), we divide the PL measured when focused optimally on a pillar by that obtained when focused on the unpatterned surface (figure \ref{PL_enhancement}\,c). We find that there is a maximal increase of roughly 11.5 times in the measured PL intensity from NV ensembles contained in nanopillars, which is in agreement with prior studies on the coupling of single NV centres to similar tapered nanopillars (showing a roughly 10$\times$ increase in PL) \cite{2543809120150114,4837081120100301}. However, in contrast to single NV centres the high density of our implanted layer results in uniform emission intensity across the whole array, whereas significant brightness variation from pillar to pillar often occurs in the single or few emitter regime due to implantation statistics and random radial placement of emitters within the pillar \cite{doi:10.1063/1.4871580,4837081120100301}. We find that the range of diameters from 700\,nm to 900\,nm is optimal for maximizing PL from single pillars (within the range of diameters tested), but note that the drop in collection for the 400 and 500\,nm pillars could be a result of fabrication defects related to the resolution of our EBL system. Regardless, the minimal variation in PL across the 700-900\,nm range of diameters implies that pillars in this range possess considerable tolerance to fabrication irregularities. \newline

\noindent PL spectra were measured from each pillar array and the control area using 1\,mW CW excitation at 532\,nm, where both excitation and collection were performed from the top of the sample (ie. not through the pillars) to minimize changes to the spectra due to wavelength-dependent collection efficiencies in the pillars \cite{2543809120150114}. We observe a decreased relative contribution to the spectral region corresponding to the neutrally-charged NV centre (NV\textsuperscript{0}) (roughly 575\,nm to 640\,nm) on the pillars compared to the flat surface, (figure \ref{PL_enhancement}\,d). To quantify this observation, we utilise a spectral decomposition technique based on non-negative matrix factorisation (NNMF) to find the two most-likely component spectra that make up our entire data set (figure \ref{PL_enhancement} e) \cite{Orth:18,BERRY2007155}. As expected, these components closely resemble the emission spectra of the negative and neutrally-charged NV centres \cite{000314343500001n.d.}. Note that a small amount of mixing between these two components is apparent and results from the number of spectra fed into the NNMF algorithm (82 total) - an increased number of training spectra would yield improved estimates. Using these component spectra, we then estimate the fraction of negatively-charged NV centres that contribute to each measured spectrum by scaling the (integrated) spectral contribution of each component by the excited-state lifetimes of their respective charge-states, taken here to be 12\,ns for NV\textsuperscript{-} and 16\,ns for NV\textsuperscript{0} (figure \ref{PL_enhancement} f) \cite{PhysRevA.97.063408}. We find 78$\pm$2\% of fluorescent NV centres on the unpatterned surface in the NV\textsuperscript{-} state, consistent with prior observations under 532\,nm excitation for single NV centres \cite{000314343500001n.d.}, but observe a steadily increasing proportion of NV\textsuperscript{-} as tip diameter is decreased.\newline

\noindent The change in the NV charge state (as seen in the collected PL) in the pillar arrays may be explained by one of the following effects: (i) An increase in local laser excitation power due to reduced reflection from the pillar tips compared to the unpatterned surface; (ii) wavelength-dependent emission patterns due to in-plane grating effects caused by neighbouring pillars \cite{Li2015}, which may change the apparent ratio of NV\textsuperscript{-} to NV\textsuperscript{0}; (iii) surface band-bending effects, though one would expect more NV\textsuperscript{0} with an increased surface-to-volume ratio \cite{Broadway2018}; (iv) constrained diffusion of negative charge in the pillars, as electrons liberated from photo-excited ensembles of NVs and nitrogen impurities can diffuse over substantial distances and may affect NV charge-state ratios \cite{Jayakumar2016}. Regardless of the origin of the effect, an increase in the contribution of the NV\textsuperscript{-} spectrum to the total emission spectrum will increase the spin contrast in NV\textsuperscript{-}-based sensing, while a true increase in the NV\textsuperscript{-} population will also increase the number of defects available to interact with the environment. Thus, the pillar arrays offer an additional enhancement in measurement sensitivity via the parameters $C$ and $N_{NV}$ in Eq. 1, on top of the increase in collection efficiency ($\epsilon$).

\begin{figure}
\includegraphics[width = \columnwidth]{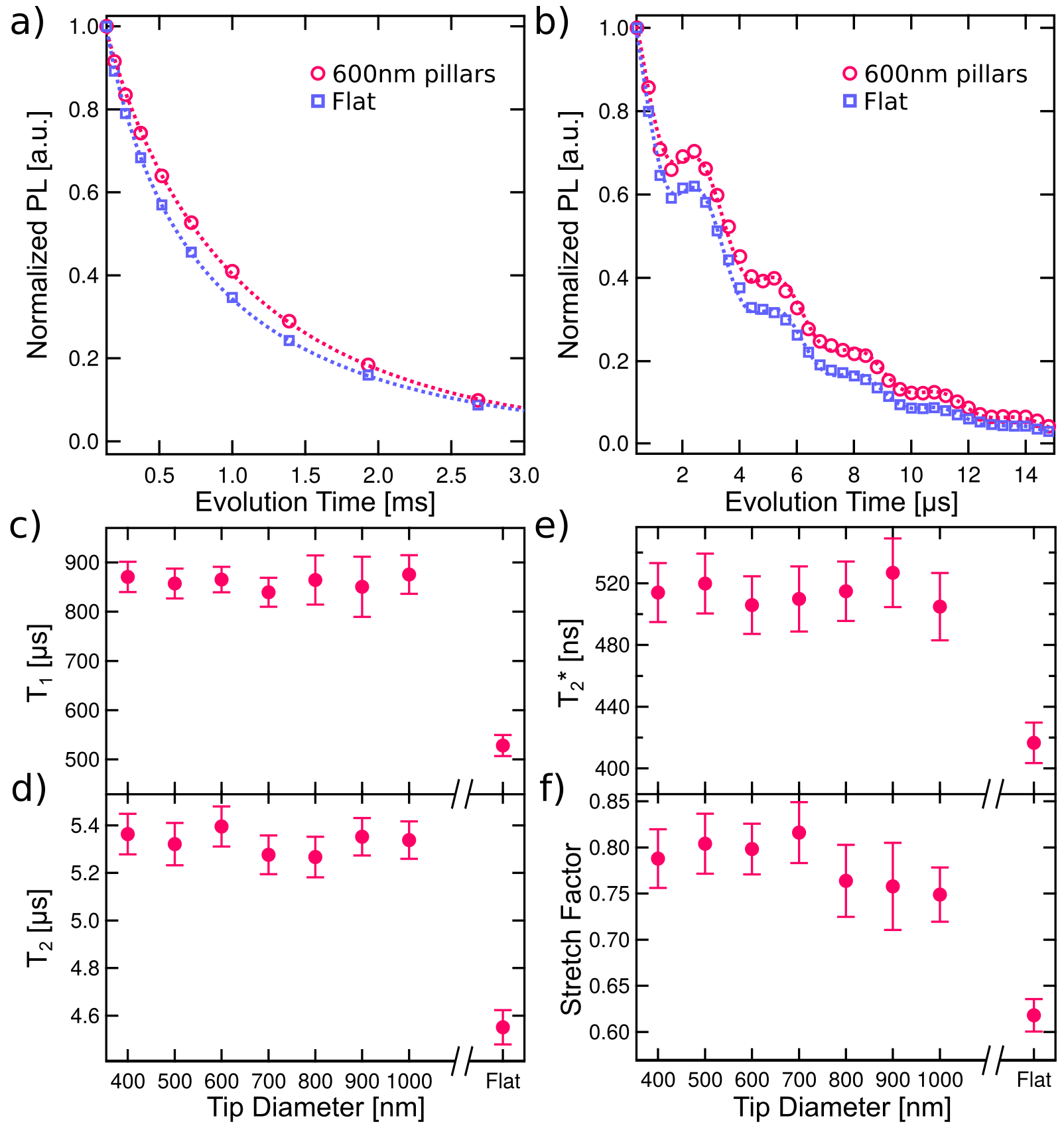}
\caption{ \textbf{Characterisation of spin properties of pillar-hosted ensemble NV. More than 500 pillars were investigated for each diameter by widefield measurement.} a) Representative spin-relaxation (T\textsubscript{1}) decay curves taken from an array of 600\,nm pillars and an unstructured surface. Dashed lines are stretched exponential fits. b) Representative spin coherence (T\textsubscript{2}) curves measured using a Hahn-echo pulse sequence for an array of 600nm pillars and the flat surface. Dashed lines are exponential decay fits modulated by a cosine function (see text). c) T\textsubscript{1} times measured for all fabricated pillar diameters and the flat surface. d) Exponential stretch parameters obtained from fits to T\textsubscript{1} decay curves. The increased values in the pillars reflect higher-frequency local magnetic noise. e) T\textsubscript{2} decay times measured for the different pillar diameters and the flat surface. f) T\textsubscript{2}* decay times for the different pillars deduced from high-resolution ODMR spectra. Error bars are fit uncertainties in all cases.}
\label{spin_properties}
\end{figure}
\section*{Spin Characterisation}
\noindent To investigate the impact of the fabrication process on the spin properties of the NV centres, widefield measurements of spin relaxation (T\textsubscript{1}) and coherence (Hahn-echo T\textsubscript{2}) times were performed using a home-built widefield imaging setup (figure \ref{overall_diagram}\,a) \cite{Simpson2016}. T\textsubscript{1} decay curves were fit using exponential decay functions of the form $PL(\tau)\propto \exp(-(\frac{\tau}{T_{1}})^s)$, where $\tau$ is the evolution time and $s$ is a parameter ranging from 0.5 to 1 that accounts for the distribution of T\textsubscript{1} times from the NV ensemble due to the local magnetic noise environment \cite{Hall2016}. T\textsubscript{2} decay curves were fit using a modulated exponential decay function of the form $PL(\tau)\propto ($A$-$B$\, \cos{(\omega \tau + \phi))} \exp(-\frac{\tau}{T_{2}})$, where A, B, $\omega$, and $\phi$ are fit to best match the coherence revivals due to native \textsuperscript{13}C in the sample (figure \ref{spin_properties} b). T\textsubscript{2}* was measured from high-resolution ODMR spectra by taking the FWHM of N\textsuperscript{15} hyperfine interaction peaks \cite{PhysRevB.97.085402}.\newline

\noindent The nanopillars improved NV\textsuperscript{-} T\textsubscript{1} by approximately 60\% and T\textsubscript{2} and T\textsubscript{2}* times by 20\% compared to the unpatterned surface for all pillar diameters tested (figure \ref{spin_properties}\,c - e). This ensemble behaviour contrasts the typical observation that the spin properties of single NV centres are deteriorated (or, at the very least, preserved) by the fabrication of nanophotonic structures \cite{2543809120150114,doi:10.1063/1.4871580}. As the pillars are expected to increase excitation intensity by focusing, we do not attribute the change in T\textsubscript{1} to inter-defect electron tunneling during dark intervals, which should lead to shorter measured T\textsubscript{1} times as excitation power increases \cite{BluvsteinChargeState}. Interestingly, the stretch parameter used in the T\textsubscript{1} fits shows a similar dependence to the coherence times (figure \ref{spin_properties} f), which implies a higher-frequency magnetic noise environment for pillar-hosted NVs versus the unstructured surface \cite{Hall2016}. We therefore suspect that the coherence improvements result from a modification of the charge populations and dynamics of the near-surface defects in the pillars, which are the primary sources of noise that limit dense (near-surface) ensemble NV coherence times \cite{PhysRevB.97.085402}. Using both the spin and photoluminescence enhancements reported here, the expected improvement in ODMR sensitivity offered by the pillars as per Eq. 1 is calculated to be 3.8$\pm$0.25 for the 800\,nm (brightest) pillars.\newline
\begin{figure*}
\includegraphics[width = \textwidth]{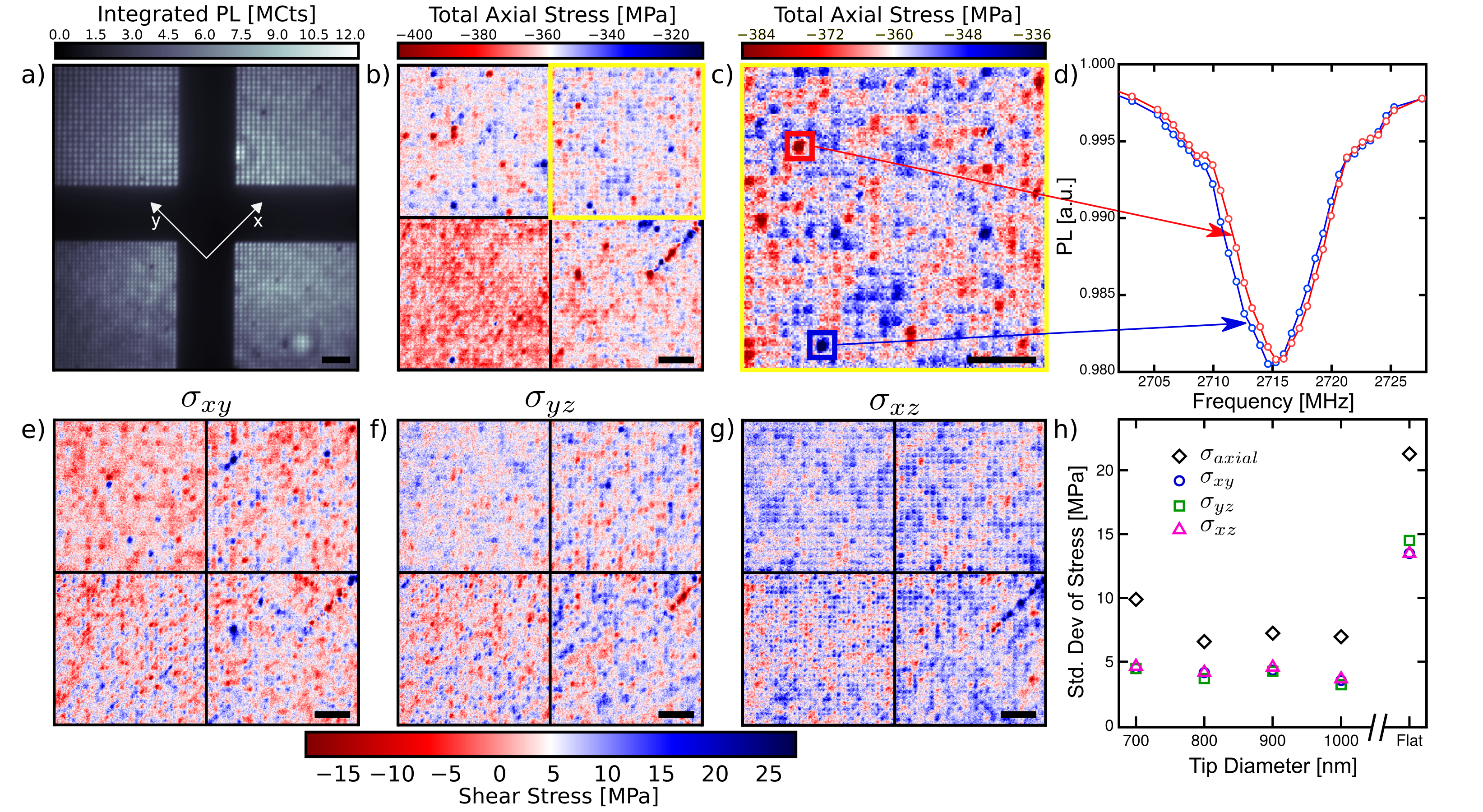}
\caption{\textbf{Mapping of mechanical strain in nanopillar arrays.} a) Widefield PL image showing four arrays of fluorescent nanopillars with different diameters (clockwise from top-left, pillar diameters are 1000\,nm, 900\,nm, 800\,nm, and 700\,nm). Also shown is the $xy$ coordinate system used to define the stress tensor components. b) Maps of total axial stress for the four pillar arrays. A polishing defect is visible in the 800\,nm array (bottom right). c) Zoomed image of axial stress on the 900\,nm pillar array, with two representative pillars highlighted. d) Zoom of one of the 8 ODMR peaks for the two pillars highlighted in c), showing a shift in one of the NV families of roughly 1\,MHz due to strain. e), f), g) maps of shear stress tensor components $\sigma_{xy}$, $\sigma_{yz}$, and $\sigma_{xz}$ respectively. h) Standard deviations in strain measurements across the measured arrays and the unpatterned surface. As these values are indicative of measurement noise, an improvement in shear stress sensitivity $>$ 3 is apparent for all pillar diameters, while there is an improvement in sensitivity to axial stress of around 2. The scale bar is 10\,\micro m in all instances.}
\label{stress_maps}
\end{figure*}
\section*{Stress Tensor Imaging}
\noindent The closely-packed nature of the pillar arrays lends them applicability as a widefield imaging tool. By performing ODMR spectral imaging (ie. acquiring a spectrum for each pixel in an image) it is possible to obtain two-dimensional images of quantities such as magnetic field, electric field, and mechanical strain \cite{Tetiennee1602429, Broadway2018, StressMappingPaper}. We illustrate this capability on the nanostructured surface by forming images of the stress tensor components on our brightest pillars utilizing the method introduced in Ref. 36. Shown in figure \ref{stress_maps}\,(a) is a widefield PL image of the area we studied, consisting of arrays with pillar diameters of 1000nm, 900nm, 800nm, and 700nm (clockwise from top-left), and the resulting spatial maps of total axial stress ($\sigma_{\rm axial}=\sum \sigma_{ii}$) in these pillars are displayed in figure \ref{stress_maps}\,(b) and (c). An example of an ODMR shift of a particular NV orientation from two representative pillars is displayed in panel (d), which shows an $\approx$2\,MHz shift between the two (note: to extract the full stress tensor we fit all eight ODMR peaks to the NV spin Hamiltonian as outlined in ref. \cite{StressMappingPaper}).\newline 

\noindent We find a steady increase in the magnitude of total axial stress as the pillar diameter is decreased. Conversely, the three shear stress components $\sigma_{xy}$, $\sigma_{yz}$, and $\sigma_{xz}$ (figure \ref{stress_maps}\,e - g) show no dependence on pillar diameter but do trend from more negative values in $\sigma_{xy}$ to more positive values in $\sigma_{xz}$. Note also that the magnitudes of total axial as well as the three components of shear stress are comparable to those measured for the unpatterned surface, which indicates that the fabrication process has had a minimal impact on in-built mechanical strain. We find that the average measurement noise is lower for the pillar arrays than for the unpatterned surface, as evidenced by a lower standard deviation in all measured components of stress (figure \ref{stress_maps}\,h). Since the measurement times for the pillars and the unstructured surface are the same, we infer an increase in ODMR specral imaging sensitivity for the different components $\sigma_{axial}$, $\sigma_{xy}$, $\sigma_{yz}$, and $\sigma_{xz}$ of 2.8, 3.3, 3.7, and 3.2, respectively. These values are consistent with the sensitivity increases of around 3.8 predicted by equation 1, though we suspect the somewhat lower values result from the fact that there is a clear variation in stress from one pillar to the next, so the standard deviations across the pillar arrays are not purely representative of the measurement noise floor. 
\section*{Conclusion}
\noindent In conclusion, we have fabricated and characterised closely-packed arrays of diamond nanopillar waveguides containing discrete, dense ensembles of NV centres. The pillars showed more than an order-of-magnitude improvement (11.5 times) in collected PL intensity over the unpatterned surface, with little variation over the optimal range of diameters from 700-900\,nm. The fraction of the PL spectrum belonging to the negatively-charged NV centre was also found to increase with decreasing pillar diameter, and we found average increases in T\textsubscript{1} of 60\% and in Hahn-echo T\textsubscript{2} and T\textsubscript{2}* of 20\% for pillar-hosted ensembles over unpatterned surfaces. These improvements indicate that an increase in widefield measurement sensitivity of at least 3.7 times can be achieved through nanostructuring of the diamond surface. This comes at the cost of spatial resolution being limited by the array pitch, but the apparent lack of pillar size/spacing dependence on coherence times suggests that similar results would be obtained with pitches below the diffraction limit of NV\textsubscript{-} emissions (100-200\,nm), which would eliminate this drawback. Finally, we demonstrated widefield imaging of the intrinsic stress in the pillar tips, showing that implanted ensemble NV can be used to image the stress tensor in diamond micro-and-nanomechanical devices. Our results demonstrate a simple method for increasing the sensitivity of widefield measurements with nitrogen-vacancy ensembles, thereby reducing acquisition times and opening up possibilities for higher-speed imaging of dynamic magnetic, electrical, thermal, and mechanical processes at the micro-and-nano scales.

\section*{Acknowledgments}
We acknowledge support from the Australian Research Council (ARC) through grants DE170100129, CE170100012, and FL130100119. D.J.M. and D.A.B. are supported by an Australian Government Research Training Program Scholarship. This work was performed in part at the Melbourne Centre for Nanofabrication (MCN) in the Victorian Node of the Australian National Fabrication Facility (ANFF).

\bibliography{pillars_charac_paper_bib}

\end{document}